# A SYSTEMS-BASED APPROACH TO MULTISCALE COMPUTATION: EQUATION-FREE DETECTION OF COARSE-GRAINED BIFURCATIONS

C. I. Siettos[1], R. Rico-Martinez[2] and I. G. Kevrekidis[3],*,
[1]School of Applied Mathematics and Physical Sciences, National Technical University of Athens, 9, Heroon Polytechniou St., Zografou, GR 157 80, Athens, Greece
[2]Instituto Tecnologico de Celaya, Depto. De Ingenieria Quimica, Celaya, Gto. 38010 Mexico
[3]Dept. of Chemical Engineering, Program in Computational and Applied Mathematics and Dept. of Mathematics, Princeton University, Princeton, NJ 08544, USA

*Abstract*

We discuss certain basic features of the *equation-free (EF) approach* to modeling and computation for complex/multiscale systems. We focus on links between the equation-free approach and tools from systems and control theory (design of experiments, data analysis, estimation, identification and feedback). As our illustrative example, we choose a specific numerical task (the detection of stability boundaries in parameter space) for stochastic models of two simplified heterogeneous catalytic reaction mechanisms. In the equation-free framework the stochastic simulator is treated *as an experiment* (albeit a computational one). Short bursts of fine scale simulation (short computational experiments) are designed, executed, and their results processed and fed back to the process, in integrated protocols aimed at performing the particular coarse-grained task (the detection of a macroscopic instability). Two distinct approaches are presented; one is a direct translation of our previous protocol for adaptive detection of instabilities in laboratory experiments (Rico-Martinez et al., 2003); the second approach is motivated from numerical bifurcation algorithms for critical point detection. A comparison of the two approaches brings forth a key feature of equation-free computation: computational experiments can be easily *initialized at will*, in contrast to laboratory ones.

**Introduction**

The purpose of this paper is to articulate certain connections we perceive between the recently developed equation-free approach to modeling and computation of complex/multiscale systems on the one hand, and fundamental tools of systems and control theory and practice on the other. We will also illustrate these connections through simple examples. Even though these connections are straightforward, our experience is that they open up some truly interesting, and, we believe, powerful alternative avenues to the use of microscopic/stochastic simulators.

Most engineering systems we try to model are inherently multiscale (consider, as an example, the macroscopic flow of a Newtonian fluid, arising through molecular interactions). Typically, however, and for practical modeling purposes, we are not interested in the detailed evolution of the system state (the positions and velocities of each molecule). Instead, we are interested in the dynamic evolution of certain coarse-grained *observables* ("outputs" of the molecular system) such as macroscopic density, momentum and energy fields. These observables are often low order moments of the evolving molecular distribution. In many cases of engineering interest (e.g. fluid mechanics) it is possible to write accurate macroscopic evolution equations that close at the level of the quantities of interest (e.g. the Navier-Stokes equations). In non-Newtonian cases, additional moments (e.g. stresses) have to be included as observables (new dependent variables), and larger, more complex sets of macroscopic evolution equations must be written. This is clearly a *model reduction* process: we go from models like molecular dynamics, with enormous numbers of degrees of freedom, that are "correct" but extremely difficult to use, to more useful and economical models (ODEs, PDEs, DAEs, PDAEs) in terms of a few dependent variables.

---

* To whom all correspondence should be addressed

The essential reduction process, which is almost always based on separation of time scales (and concomitant space scales), is the derivation of *closures:* formulas that embody the effect of unmodeled scales and variables on the ones we retain. A celebrated such closure for Newtonian fluid mechanics is Newton's law of viscosity, that expresses stresses (in effect, higher order moments of the molecular distribution, which evolve on a faster time scale) as a functional of the velocity field (in effect, lower-order, slower evolving moments). These closures may be derived experimentally, from extensive observations, often in the form of correlations, or they may be derived mathematically from the microscopic models themselves under appropriate assumptions (such as the Chapman-Enskog expansion of the Boltzmann equation (Chapman and Cowling, 1964)). Traditional engineering modeling starts with obtaining such closures, and writing the corresponding macroscopic evolution equations in terms of the relevant, coarse-grained system observables ("outputs"). When reasonably accurate macroscopic equations have been explicitly constructed, the mathematical and computational tools developed for the study of such equations are brought to bear on their analysis. Simulation, stability and bifurcation analysis, optimization, controller design: the development and implementation of algorithms for performing such tasks on traditional evolution equations constitute a large component of engineering education and practice.

When accurate macroscopic evolution equations for a system of interest can be derived, this "traditional path" of computer-assisted modeling (first obtain governing equation through closures, then study and analyze the equations) is "the way to go". However, in current modeling practice, the need constantly arises for studying problems for which good, quantitative closures have not yet been derived (experimentally or theoretically) and the only available modeling tool is direct simulation with a microscopic (atomistic, molecular, agent-based) code. This simulation "lives" on extremely fine space and time scales, and it is very often prohibitive; hardware evolution is not going to bridge this "scale gap" soon enough. Systematic improvements in the fine-scale algorithms (e.g. more efficient thermostats in certain MD computations) are very important, but the scale gap still remains. We have a "correct" fine-scale model, which is extremely difficult and time consuming to simulate; and accurate, closed form, macroscopic evolution equations for the coarse-grained quantities of interest are not available, nor can they be easily derived.

This is the regime that equation-free modeling tries to address: problems where a trustworthy fine scale simulator is available, *and we have reason to believe that macroscopic coarse-grained equations in terms of a number of system observables exist* in principle*, even if they are not available in closed form.* This premise (believing that a model of the system *does* close at a certain coarse-grained level, even if we do not know how to write the closure) is a fundamental premise of equation-free computation. It is also the typical situation one encounters when trying to model or control a new experimental system: thinking of the system as an input-output black box, one asks which (and how many) the relevant variables are, in terms of which one would write a deterministic input-output model. At this stage we will assume that we know which and how many these important variables are; for our laminar Newtonian isothermal flow example, this would involve density and momentum fields. We have been taught that these are "good variables" for such problems as undergraduates; yet in new problems we will not know which, nor even how many, the right variables will be. Data analysis tools that address this question become therefore vital, and we will return to this issue in our Discussion section.

Consider how an accurate, macroscopic evolution equation is used in scientific computation: its use depends on what numerical task we want to perform. For dynamic simulation we need to repeatedly call a subroutine that, given the system state, returns the *time derivative* (evaluates the right-hand-side of the evolution equations). For implicit integration, we will also need a subroutine that evaluates the linearization (the Jacobian) of these equations. Finding steady states through a contraction mapping, such as Newton iteration, also requires repeated evaluations of the right hand side and the linearization of the model. Thinking a little further along these lines shows that in most scientific computing tasks, the *actual form* of the macroscopic equation is *never used* – we only need *a few terms of its Taylor series,* usually the zeroth and first order terms, at most the second order terms, evaluated at a *sequence of state instances.* These instances are not known *a priori* – the second Newton iterate depends on the first, and so on. (A case where this is *not* true, i.e. where the functional form of the equation *is* required, occurs in global optimization, where one computes under-estimators of various terms in the right-hand-side of an equation).

Depending on the task we wish to perform, the "main algorithm" calls the subroutine containing the model and obtains "on demand" the right-hand-side (RHS) and/or its linearization at the current state; these numerical values are processed based on the task in question (different processing for numerical integration, different processing for fixed point computation, different processing for optimal control computations) and, based on this processing, *new state instances* are arrived at for which we want to call the model subroutine again. It is clear that scientific computation does not require *a closed model formula* – it only requires a couple of terms of the Taylor series of the model, but it requires them at many points, and these points are not known at the beginning of the computation. Whether the model subroutine contains a compact closed formula that can fit in a single programming line, or whether it contains an enormous lookup table, the "main algorithm" does not know and does not care: it requests certain numerical quantities to compute with, receives them, and proceeds in identical fashion.

The simple (but obvious) next step, is to consider the case in which the "model subroutine" is neither a one-line compact formula, nor a precompiled lookup table – instead, it consists of a request to a laboratory experimentalist for the first couple of terms of the Taylor expansion of the system RHS *at a prescribed state*. It should be conceptually clear that, *if it is easy to initialize a physical experiment at a given state*, then a few experiments can be performed from which the relevant numbers can be estimated and returned to the "main algorithm" – in which case one is performing numerical integration, fixed point computation, or in principle any other numerical task using *not a model,* but the physical experiment itself, *appropriately initialized*.

This line of thinking shows that, in principle, one can perform many scientific computing tasks *without a model*, as long as one has access to the physical experiment itself, *and can initialize it at will*. Short experiments, at states prescribed by the "main algorithm" give values for residuals and Jacobians *at the state point of interest;* the "main algorithm" processes them and suggests a new state point at which similar information is required – and the process continues until the result of interest is arrived at. In effect, traditional numerical algorithms become *protocols for the design of experiments* (physical or computational) so that we arrive at the final desired result.

Clearly, many elements of systems theory are playing a role in this approach: which experiments do we need to perform in order to estimate the local linearization of a model? How should we process the (probably noisy) results to get our best estimate of these quantities? Is a local linear model consistent with our data? Yet the one (small but vital) issue is that one requires this information *many times* and *at prescribed state points*. Here arises, in our opinion, the crucial difference between laboratory experimentation, and *computational experimentation*: it is almost trivially easy to initialize a computational experiment "at will" at several, different, prescribed initial conditions – one only changes the initial data passed in the subroutine call. On the other hand, it is typically extremely difficult, or even practically impossible, to initialize a laboratory flow at several, different, prescribed velocity fields. This is the reason (or at least *one* important reason) why such approaches are not typically used in laboratory experimentation.

In equation-free computation we do not have a subroutine containing the macroscopic, coarse-grained model. But we do have a subroutine that can evolve the fine scale model, and we in principle *can* initialize the fine scale model consistently with values of its macroscopic observables (construct distributions consistent with a few of their lower moments). This operation (*fine scale* initial conditions from *coarse scale* observables) is called *lifting* in EF terms; it is a one-to-many operation (there are clearly many ways to initialize distributions with the same mean and variance!) and we will return to its discussion below. When the "main code" asks for specific Taylor series information of the unavailable, coarse-grained model at a particular state point, we then "simply" initialize the fine scale code consistently, perform several short bursts of computational experimentation with it, and process the results to estimate the numbers of interest – these numbers are then *fed back* to the "main code" which can continue its scientific computation. The procedure occurs again and again, until the final result is reached. It is reached without ever finding a closed form macroscopic model; yet it is reached using the fact (the belief) that such a model exists, that we know in terms of which variables it closes, and *having access to the fine scale code.* We are literally doing coarse-grained scientific computation directly on the fine-scale model, without the intermediate of a closed formula – computational experimentation and estimation are the tools we used to substitute function and Jacobian evaluations. It is important to reiterate that we must be able to initialize the fine scale code *easily* – at fine scale states consistent with specific values of the coarse-grained observables.

What does one benefit by running the fine scale code this way? Normally, one would initialize the fine scale code (e.g. the molecular dynamics), prescribe parameter values, let the code evolve over long times, and observe its output. This way we are simulating on the computer temporal evolution as it would occur in nature – we are doing on the computer our best approximation of what we would see in the laboratory. What if the information we need, however, is not the temporal evolution itself? What if, for example, we need to find stationary states, and we are only simulating until we get eventually close to a stationary state? It is in these cases that equation-free computation may provide some computational benefits. Steady states are often found much faster from fixed point computation – rather than direct simulation (especially unstable ones that cannot in principle be found by direct simulation). In performing Newton iterations, one uses the same model (the same equations) as when one is integrating in time – but uses them *differently*. Fixed point algorithms *take advantage of the local smoothness* in the neighborhood of a steady state to converge on it with a relatively much smaller effort than long, protracted integration. (An important caveat is that these potential computational savings are strongly problem dependent, and that Newton iteration requires good initialization!) It is smoothness and Taylor series in the neighborhood of the steady state that allows us this computational economy in reaching the same result (the steady state) by Newton rather than by direct simulation.

Consider now a steady state of the laminar Navier-Stokes equations (to revisit our discussion example). It is the density and momentum fields that are at steady state – the molecules are not; they keep evolving and colliding. The Jacobian of the molecular dynamics equations cannot be used to help us converge to the coarse-grained steady state – we need derivatives with respect to the coarse-grained variables themselves. It is these coarse-grained derivatives (derivatives in time, derivatives with respect to parameters, derivatives with respect to coarse-grained variables) that have the potential for overall computational savings – these quantities allow us to "jump in time" in a forward Euler step; to "jump in phase space" in a Newton step or in a conjugate gradient step; to "jump in parameter space" in a continuation step, etc. Since we have no coarse-grained equations, we cannot differentiate them to obtain coarse-grained derivatives – we have to *estimate* the coarse-grained derivatives from short bursts of appropriately initialized microscopic simulation, in order to link to continuum numerical analysis algorithms and possibly accelerate our computational time-to-results.

In summary, equation-free computation is an "input-output" approach to modeling complex or multiscale systems. When a fine scale code is available, and we have reason to believe that coarse-grained equations can in principle be derived, we do not need to derive them to compute with them. We can perform many scientific computing tasks with them by calling the fine scale solver as *a computational experiment* that can be *initialized at will*. Short bursts of appropriately initialized computation with the fine scale code allows for the *on demand* estimation of the numerical quantities required for coarse-grained scientific computation. Elements of systems and control theory (design of experiments, estimation, variance reduction, model reduction, model identification) become a vital component of the EF process, in linking the fine scale simulation with the coarse-grained computational tasks. To illustrate these points, we will choose a particular type of a fine scale simulator (the Gillespie stochastic simulation algorithm, SSA (Gillespie, 1976, 1977)). We will also choose one particular type of scientific computing task (the computation of coarse-grained bifurcation points).

Bifurcation points (such as turning points) often mark the boundary between stability and instability for a dynamical system. Bifurcation theory is well established and used for the stability and parametric analysis of nonlinear continuum system models ranging form ecology to materials science. In many cases it is required that the system of interest operate at open-loop unstable equilibria or even at marginally stable (critical) points. Presuming modal controllability and the accurate knowledge of the location of these critical points, an arsenal of control design techniques can be used to cope with system stabilization there. Several methods have been also proposed to deal with the stabilization of nonlinear systems in the case where the critical modes are uncontrollable for the corresponding linearized system: sufficient conditions for the local feedback stabilization and control of static and Hopf bifurcations of nonlinear systems are given in (Abed and Fu, 1986a, 1986b; Gu et al., 2000) while bifurcation control, based on normal forms, via state feedback with a single uncontrollable mode is addressed in (Kang, 2000). The incorporation of washout filters (Lee, 1991; Wang et al., 1999) is used to cope with system stabilization while preserving the equilibria of the original system in the presence of model structural uncertainties. Global stabilization of bifurcations, using Lyapunov tools, changing the type of bifurcation from subcritical to supercritical is presented in (Krstić et al., 1998, 2000). Anderson et al., (1999) and Rico-Martinez, et al. (2003), proposed an adaptive control method for the detection of instabilities in experimental systems, while a robust method for experimental bifurcation analysis for periodically driven systems is addressed in (Langer and Parlitz, 2002). All these approaches operate at the continuum/macroscopic level, possibly accounting for a degree of uncertainty.

We are interested in extending these techniques so that they can be used in problems for which coarse-grained evolution equation models are not available. The task we want to perform is at the coarse-grained, systems level, while the available simulator is an atomistic (in this case, stochastic) one, which cannot be direcly coupled with continuum level tools. Over the last few years, it has been established (Gear et al., 2002; Makeev et al., 2002; Runborg et al., 2002; Siettos et al., 2003a; Kevrekidis et al., 2003) that coarse timesteppers can serve as an equation-free bridge between microscopic problem descriptions and continuum level computational methods. Based on the concept of coarse timesteppers we address the development of an integrated framework that enables miscroscopic simulators, under certain assumptions imposed by the criticalities sought, to converge to their open-loop coarse bifurcation points. We will present two different approaches: the first one is directly based on the approach presented in (Rico-Martinez, et al., 2003). An adaptive control scheme was used in that work to drive a laboratory experiment to its critical point and keep it there; here we drive a microscopic simulator (again an experiment, but now a computational one) to its coarse critical point and keep it there. The steering of the microscopic timestepper is achieved through the manipulation of the bifurcation parameter, which becomes our control variable. The second approach is motivated by numerical bifurcation theory and incorporates a direct method for the detection of bifurcation points in the equation free framework. In this case, several brief computational experiments are designed based on the numerical quantities required in the bifurcation codes; these experiments are executed and discarded after the relevant quantities are estimated, and new ones are initialized subsequently. It is worth noting that our "input-output" algorithms are applicable for both "legacy" black-box continuum dynamic simulators as well as for microscopic/ stochastic simulators (the application of choice here). The two approaches may also complement each other; for example, the adaptive control-based detection protocol can be used to provide a good initial guess of the critical point to the numerically-motivated algorithm.

The paper is organized as follows: in the next section we briefly describe the concept of coarse timesteppers (Gear et al., 2002, Makeev et al., 2002, Kevrekidis et al., 2003). We then present our two alternative approaches. To illustrate our methodology we used stochastic simulation of two simple surface reaction mechanisms: the first mechanism results from a simplification of the kinetics of NO reduction by $H_2$ on Pt and Rh surfaces. The second one is a simplified kinetic description of the CO oxidation reaction with an inert site-blocking, reversibly adsorbing species; solution branches of both mechanisms possess turning point bifurcations. We will conclude with a brief discussion.

**Coarse Timesteppers and Detection of Open-Loop Coarse Bifurcation Points for Microscopic Simulators**

*The concept of coarse timesteppers*

The coarse timestepper is a fundamental building block of equation-free algorithms – it is the way to obtain macroscopic input-output information from a microscopic simulator. The main assumption behind the coarse timestepper (as discussed in some detail above) is that a coarse-grained model for the fine-scale dynamics exists and closes in terms of a few coarse-grained variables (observables); and that we know a good set of such observables. Typically these are low-order moments of microscopically evolving distributions, and the existence of a coarse-grained model implies that the higher order moments of the distributions become, relatively quickly over the time scales of interest, "slaved" to the lower, few, "master" ones. What a coarse timestepper does, in fact, is providing a closure on demand ("just in time" (Cybenko, 1996)); relatively short bursts of fine scale simulation naturally establish this slaving relation (refer to Gear et al., 2002; Makeev et al., 2002; Siettos, et al., 2003a; Kevrekidis et al., 2003) for more detailed discussions). Briefly, once the appropriate macroscopic observables have been identified, the coarse timestepper consists of the following essential components (see Kevrekidis et al., 2003; Makeev et al., 2002; Runborg et al., 2002):

(a) Selecting values of the coarse initial conditions (I.C.'s);
(b) Using a *lifting* operator it to transform the coarse I.C.'s to one or more consistent microscopic initial conditions;
(c) Evolving these microscopic distribution(s) through the fine scale simulator for a short macroscopic time *T*- and
(d) *Restricting* the resulting microscopic distribution(s) back to their moments at the reporting horizon time *T* (i.e., finding values of the observables after time *T*).

The above "black box" coarse timestepper is a mapping $\Phi_T$:

$$x_{k+1} = \Phi_T(x_k, p_k) \quad \Phi_T: R^m \times R^q \rightarrow R^m, \qquad (1)$$

where the vector $x$ denotes the coarse variables and $p \in R^q$ denotes the vector of system's parameters.

Computational coarse timestepping circumvents the derivation of explicit macroscopic equations, (i.e. the analytical derivation of $\Phi_T$), while it is still able to deliver systems level information operating directly on the microscopic evolution rules: e.g. coarse (macroscopic) steady states can be obtained as fixed point, using *T* as sampling time, of the mapping $\Phi_T$: $x - \Phi_T(x, p) = 0$. The choice of *T* is associated with the (estimated) spectral gap of the linearization of the unavailable closed macroscopic equations (Siettos et al., 2003a). Calls to the coarse timestepper from *coarsely nearby* initial conditions allow the estimation of the action of the linearization of the map $\Phi_T$ on known vectors, since

$$D\Phi_T(x) \cdot v \approx \frac{\Phi_T(x + \varepsilon v) - \Phi_T(x)}{\varepsilon} \qquad (2)$$

Using the coarse timestepper to evaluate the "time-T map" of the coarse-grained model, as well as the action of this map's Jacobian on prescribed vectors, underpins the equation free computational framework; it serves as a bridge between microscopic / stochastic simulators and macroscopic tasks such as coarse bifurcation analysis, control design and optimization calculations (Theodoropoulos et al., 2000; Kevrekidis, et al., 2003; Makeev et al., 2002; Runborg et al., 2002; Gear et al., 2002; Siettos et al., 2003a, 2003b; Armaou et al., 2004, Armaou et al., 2005) Many other computational approaches to coarse-graining microscopic simulations for control purposes have been proposed (see for example Gallivan et al., 2001; Ni and Christofides, 2005; Lou and Christofides, 2005).

*The adaptive control-based detection approach*

As a representative system-level computational task that one might want to perform with a fine-scale model, we will select the detection of a coarse instability boundary (a coarse bifurcation point). Here we use the fine scale simulation *as an evolving experiment*, and discuss an adaptive control protocol, which, when wrapped around an existing microscopic simulator, enables it to seek such open-loop coarse bifurcation points and converge there. The protocol was originally designed for, and applied to, the on-line detection of bifurcations in experiments (Rico-Martinez et al., 2003). We assume that we know the appropriate coarse variables *x*. The fine scale simulator has been initialized and it is evolving; at any given moment in time we have access to its fine-scale state, and so we can easily observe (compute the values of) the coarse variables through a *restriction* step. We will start, for convenience, in the neighborhood of a stable coarse steady state ($x^*, p^*$). The procedure makes use of several standard systems and control theory tools as follows:

*Do Until convergence to the critical point {k=0, 1,…*

1. Implement, around the coarse observations of the evolving fine scale simulator, standard recursive system identification algorithms, such as recursive least squares and Kalman filter approaches (Astrom, and Wittenmark, 1995; Ljung, 2001), to estimate the parameters of a NARMAX coarse local model of the form

$$x_{k+1} = F_T(x_k, p_k). \quad (3)$$

The structure of this local model, e.g. a truncated, low order Taylor series, should be capable of efficiently approximating the *local*, expected, coarse dynamics of the input-output map $\Phi_T$, in a neighborhood of the current coarse equilibrium point ($x^*, p^*$). The parameter $p$ will be our control variable.

2. With such a suitable local parametric model in hand, we estimate the location (in coarse phase and parameter space) of a nearby bifurcation point: we use standard optimization techniques to find a local minimizer ($x', p'$) of the criticality condition that defines a bifurcation point. This can be stated as a constrained minimization problem of the form

$$J = \min_{x \in R^m, p \in R} \{\frac{1}{2} \vartheta(x, p)^2\}, \quad (4)$$

subject to the constraints

$$x - F_T(x, p) = 0, \quad (5a)$$

$$x^* - a < x < x^* + a, \, p^* - \beta < p < p^* + \beta \quad (5b)$$

$\vartheta(x)$ is the criticality condition; e.g. in the one-(coarse)-dimensional case, and for a turning point this becomes $\vartheta \equiv \frac{\partial F_T}{\partial x}\Big|_{x=x^*} - 1$.

A number of reasonable precautionary steps are taken: for example, at each iteration, we may use bounds in the optimization problem (eq. 5b), to ensure feasible solutions.

3. We then implement a feedback control law, to drive the *identified local model* to its estimated critical point ($x', p'$). Using the separation and certainty equivalence principles, one can design and implement linear and/or nonlinear control techniques (e.g. local stabilizing direct state-feedback controllers, wash-out filters, model predictive controllers etc.); the main point is that standard control theory tools are used around a *computational experiment*.

For simplicity, in our illustrative example we chose a discrete-time, direct state-feedback controller of the form

$$p_k = p' - K^T(x_k - x') \quad (6)$$

The values of the gain vector $K$ are calculated on the basis of a discrete linear quadratic control design, i.e. we seek a state-feedback control law that minimizes a performance criterion of the form

$$J(p) = \sum_{k=0}^{\infty} x_k^T Q x_k + p_k^T R p_k \quad (7)$$

with respect to the linearized dynamics of the identified coarse local model (3) around the estimated critical point ($x', p'$), given by

$$\Delta x(k+1) = A \Delta x(k) + B \Delta p(k) \quad (8)$$

where $A \equiv \frac{\partial F_T}{\partial x}\Big|_{x',p'}$ is the system Jacobian matrix, $B \equiv \frac{\partial F_T}{\partial p}\Big|_{x',p'}$ is the control vector.

*} End Do*

Some remarks, pertaining to the particular numerical task of critical point detection, are in order. Care should be taken in the choice of the control law with regard to the *type of bifurcation sought*, and also to account for uncertainty imposed by the approximating model. For example, while linear wash-out filters are appropriate, under certain assumptions, for the stabilization of Hopf and pitchfork-type bifurcations (Abed & Fu, 1986a, 1986b; Lee, 1991) even in the presence of model uncertainty, they are inadequate when dealing with other static bifurcations such as transcritical and saddle-node ones (Lee, 1991). Nonlinear or direct state feedback can be used, under appropriate assumptions, to stabilize the latter types of criticality (Lee, 1991; Kang, 2000). In practice, assuming system controllability, one can drive and keep the system close to the bifurcation point.

This approach turns a modeling problem (the location of a bifurcation point) into a feedback control problem *for the dynamic simulator*. Standard identification tools are wrapped around (coarse) observations of the evolving (fine scale) dynamic simulator; an objective is formulated *based on the modeling task we want to accomplish*; and standard control design tools are used to drive the (computational) experiment to satisfy the objective. Control is being performed on dynamic fine scale simulations *for coarse grained modeling purposes*. At all stages, only local, low order coarse nonlinear models are identified from fine scale system observations – we never attempt to derive global coarse-grained nonlinear models.

*The numerically-motivated approach*

In the previous section we posed a modeling objective as a control problem; and using the fine scale simulator as a computational experiment, we used standard identification and control tools to drive the simulator to satisfy the objective. Even though the process occurs on the computer, it was originally devised (and successfully used) to work on real-time laboratory experiments. Identifying a good local model is crucial in the process, and an important factor contributing to the quality of the local model is the "richness" of dynamic information provided by the computational experiment itself. Usually, to obtain rich local information from a laboratory experiment, some form of "intelligent" excitation protocol is used (as was done in obtaining the computational results below).

We now return to an important advantage of *computer experimentation*: in general, it is easy to initialize computational experiments at will (from different prescribed states). Furthermore, one does not have to continuously evolve a *single* realization of the problem on the computer – one can run *a particular* realization for a short time, and then, if different information is required, the realization can be stopped, discarded, and a different realization can be initialized. Furthermore, several realizations can be evolved in parallel. This type of freedom available in computational experimentation allows us to gather model information *not by inducing a single experiment* to visit different regions of phase space, but rather by initializing several different brief computational experiments at appropriate state space sampling locations. One can think of this as a type of importance sampling. While the tools for processing the results are the same (filtering, identification etc.), the way the information is gathered is much simpler and more direct for computational experiments. Once more, this is because computational experiments can easily be initialized at will; and because, once a certain piece of information has been gathered, they are expendable: they can be discarded, and new experiments initialized at locations "richer" towards additional information gathering.

A good example is provided by matrix-free methods of iterative linear algebra (Kelley, 1995a). One does not need to have a matrix (alternatively, identify a linear system) in order to perform certain tasks with this matrix. It is enough to have matrix-vector products for a sequence of known vectors in order to perform tasks such as solving linear equations or finding leading eigenvalues.

Matrix vector products, as we discussed above, can easily be estimated *from nearby computational experiments*, as long as the experiments can be easily initialized *at prescribed initial conditions* (the "second" initial condition is obtained from the first by adding a small multiple of the vector whose product with the explicitly unavailable matrix we want to estimate). Easy initialization makes a tremendous difference in the way we can design numerical experiments with a dynamic simulator – and allows us to use the same traditional systems theory tools (identification, controller design etc.), assembling them now in slightly different protocols.

The guiding principles for our second approach do not come from adaptive control, but rather from established numerical bifurcation algorithms. Numerical bifurcation theory provides an arsenal of tools for finding fixed points, tracing bifurcation diagrams, locating bifurcation points and tracing their loci in multiparameter space, using criticality conditions to formulate nonsingular augmented systems of algebraic equations, and building efficient contraction mappings (Keller, 1977; Abbot, 1978; Beyn and Doedel, 1981; Parker and Chua, 1989; Seydel, 1994; Govaerts, 2000; Doedel and Tuckerman, 2000). These procedures are predicated on the availability of closed form system evolution equations (i.e. "right-hand-sides" of the governing ODE or discretized PDE models) and our ability to evaluate their partial derivatives (of various low orders) with respect to parameters and/or dynamical variables. The proposed numerical-assisted framework allows us to bypass the explicit derivation of such coarse equations, and to perform numerical tasks with them only through the "intelligent" design of brief dynamic numerical experiments with the fine scale solver. A different example of such a procedure, where the objective was the construction and exploration of coarse bifurcation diagrams, can be found in our recent work (Siettos et al., 2004). There, we proposed a methodology combining "coarse timestepping" and pseudo-arclength continuation with linear dynamic feedback control, steering microscopic simulators along their coarse bifurcation diagrams and enabling them to converge on both stable and unstable open-loop coarse stationary states.

Our goal here is to extend such approaches to the problem of coarse critical point detection. To motivate our approach we briefly describe the numerical computation of steady state bifurcations using augmented equation systems. We then use the coarse timestepper to solve such augmented systems through short bursts of fine scale simulation. The "black box" coarse timestepper is again obtained through the following steps:

1. Start with an initial macroscopic guess of the coarse critical point, $(x, p)$
2. Transform the macroscopic state $x$ through a lifting operator $\mu$ to $N_{copies}$ fine, *consistent* microscopic realizations $X_{i=1,2,...,N_{copies}} = \mu x$;
3. Evolve these microscopic distributions through the microscopic simulator for a short macroscopic time $T$- and

4. Obtain the restrictions $x_{i=1,2,...,N_{copies}}(T) = M X_{i=1,2,...,N_{copies}}(T)$ for each one of the microscopic distributions; average the restrictions to get the coarse state $x(T)$.

This gives us a "black box" evaluation of $\Phi(x) = x(T)$ with $x$ as the initial condition. We now

5. Wrap around the coarse timestepper matrix-free iterative methods (quasi-Newton methods such as the Broyden, Fletcher, Goldfarb, Shanno (BFGS), direct methods such as the Nelder-Mead algorithm or Newton-GMRES methods (Kelley, 1995b)) to seek a solution ($x$, $p$, $q$) of the following system of 2m+1 equations:

$$\begin{bmatrix} x - \Phi_T(x, p) \\ (I - J(x, p))q \\ \|q\| = 1 \end{bmatrix} = 0 \quad (9)$$

where $J(x, p) \equiv \nabla_x \Phi_T(x, p)$ is the Jacobian matrix. The second equation in (9) implies that we seek critical points where the Jacobian is singular, a necessary condition for turning point bifurcations. Note that in the above formulation the explicit evaluation of the Jacobian is not required. Instead what is needed is the action of this Jacobian on vectors, which can be obtained through the timestepper using directional derivative approximations.

The convergence of this procedure does not rely on the many and strict solvability assumptions for the control problem in the first approach; given a good initial guess, it is in principle straightforward for it to locate critical (here, saddle-node, or turning) points. This "repeated initialization" numerically motivated framework may be much more practical and computationally efficient than closed-loop control-based simulation runs. A good initial guess of the critical point is of course required; conceivably, this can be provided from a previous computation of a coarse bifurcation diagram.

**Simulation results: Steering SSA models to their coarse turning points.**

We illustrate the two approaches using coarse time-steppers based on kinetic Monte Carlo models of simple surface reaction schemes. Our first illustrative example is a Gillespie "stochastic simulation algorithm" (SSA) (Gillespie, 1976, 1977) of a simplification of the mechanism for NO oxidation reaction by $H_2$ on Pt and Rh surfaces; the mean field ODE approximation for this mechanism involves adsorption, first order desorption and reaction:

$$\dot{\theta} = \alpha(1-\theta) - \gamma\theta - k_r(1-\theta)^2\theta \quad (10)$$

where $\theta$ is the coverage of adsorbed NO, $\alpha$ is the adsorption rate constant, $\gamma$ is the NO desorption rate constant, and $k_r$ is a reaction rate constant. Simulation results were obtained for $\alpha = 1$, $\gamma = 0.01$; $k_r$ was the bifurcation parameter (and, in our scheme, the control variable). The deterministic version of the model exhibits two turning points (at $k_r \approx 3.96$ and $k_r \approx 26$) as shown in the bifurcation diagram (Figure 1a). The coarse timestepper

$$\theta_{k+1} = \Phi_T(\theta_k, k_r) \quad (11)$$

of the SSA model was used as a "black box". The inset in Figure 1a gives a closer look at the region around the turning point at $k_r \approx 3.96$.

Before estimating, for the very first time, the location of the coarse turning point, we excited the fine scale simulator with a sinusoidal signal around $k_r = 6$ for a short period of time (as shown in Figure 1b), to gather data and obtain an initial local model. In figure 2a we also show the closed loop trajectories on the ($k_r, \theta$) phase plane, i.e. the open-loop bifurcation diagram. SSA simulation results were obtained using $T = 0.5$ as the reporting horizon, while the values of the number of available sites (system size), $N^2$, and the number of realizations using for averaging, $N_{run}$, were chosen here to be $800^2$ and 100, respectively. Clearly, the adaptive control based methodology succeeds in steering and keeping the SSA simulator near (actually almost on) its coarse open-loop turning point (the protocol keeps the coarse state of the microscopic simulator within a narrower region than the square symbol appearing in Figure 1a).

For the numerically motivated approach, in the estimation of matrix-vector products we used perturbations of the order of $O(10^{-2})$ in our difference approximation, declaring convergence when the residual became $O(10^{-6})$. Here the time horizon was taken to be $T=0.05$ while the number of available sites was set to $N^2=1000^2$ and the number of realizations to $N_{run}=2000$. The optimization algorithm used to solve the augmented equations was the Nelder-Mead algorithm (Kelley, 1995b). The initial guess was set to (0.3, 4.5) and the approach converged to the "correct" coarse bifurcation point.

Our second illustrative example is an SSA simulation (Gillespie, 1976 and 1977) of the reaction:

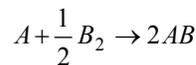

In particular we study a simplified description of the kinetics of catalytic CO oxidation, whose mean field description is given by the following ODE system:

$$\frac{d\theta_A}{dt} = a(1-\theta_A-\theta_B) - \gamma\theta_A - 4k_r\theta_A\theta_B \quad (12a)$$

$$\frac{d\theta_B}{dt} = 2\beta(1-\theta_A-\theta_B)^2 - 4k_r\theta_A\theta_B \quad (12b)$$

where $\theta_i$ represent the coverages of species (i=A,B, resp. CO, O) on the catalytic surface. The parameters $\alpha, \beta, \gamma$ are associated with CO adsorption, O dissociative adsorption and CO desorption rates, while $k_r$ is a reaction rate constant.

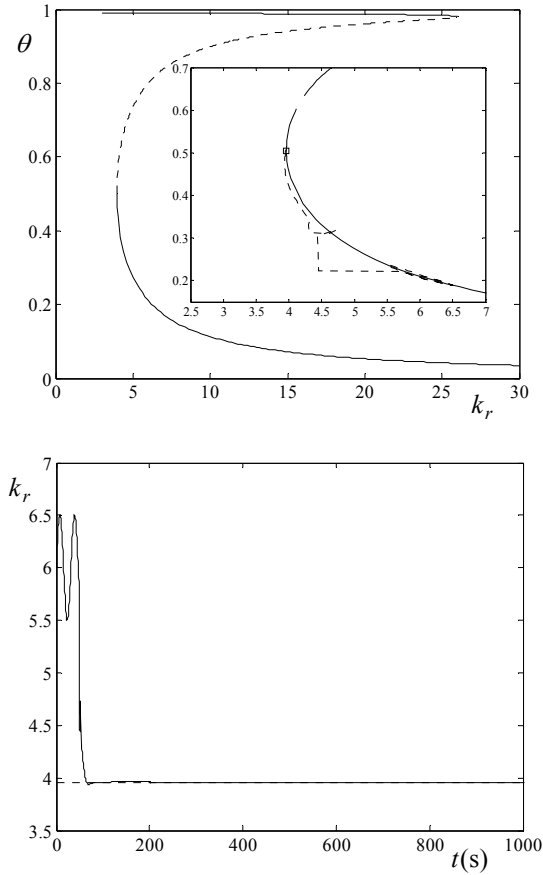

*Figure 1. (a) The bifurcation diagram for the first mean-field example. The model exhibits two turning points at $k_r \approx 3.96$ and $k_r \approx 26$. The inset shows the phase-plane trajectory (dotted line) of the closed loop response of the SSA simulator. (b)Closed loop transient of the control variable, $k_r$ (the open loop bifurcation parameter). The dashed line denotes the $k_r$ turning point value.*

Simulation results were obtained for $\alpha = 1.6$, $\gamma = 0.04$ and $k_r = 1$. The bifurcation parameter, which acted as the control variable in our adaptive control-based protocol was $\beta$.

The bifurcation diagram of the mean field model with respect to $\beta$ is given in Figure 2a. It exhibits two turning points at $\beta \approx 2.5223$ and $\beta \approx 8.4953$. Again we initially excited the fine scale simulator for a short period of time (up to 200 s), with a sinusoidal signal, to construct, through recursive estimation, an initial local model; this was then used for the first estimation of the critical point. The parameters of the model were updated at every step of the protocol using a recursive Kalman filter algorithm. The closed loop transient of the SSA simulator, obtained by implementing the adaptive control-based detection protocol, is overlaid on the open-loop bifurcation diagram. Figure 2b shows the closed-loop transient of the control variable $\beta$. The procedure drives and keep the fine scale simulator close (actually almost on) its coarse turning point (the protocol keeps the coarse state of the microscopic simulator within a narrower region than the square symbol appearing in Figure2a). SSA simulations were performed using $T=0.5$, $N^2=800^2$ and $N_{run}=100$. Starting with the initial state guess close to the turning point, the numerical bifurcation-motivated approach converged again to the "correct" coarse state and parameter values.

## Conclusions

We discussed and illustrated certain aspects of the equation-free approach to complex/mutiscale systems modeling and computation. The coarse timestepper uses the fine-scale simulation model as an experiment, which can be easily interrogated (initialized at will) to provide local information (residuals, actions of the Jacobian) of the unavailable, coarse-grained, macroscopic model. Using standard tools from control theory, one can feed these observations back to the simulator in order to steer it towards solving modeling problems (such as the automatic location of stability boundaries). Alternatively, one can use standard numerical analysis algorithms as templates to devise experimental design protocols for brief numerical experiments with the fine scale simulator that solve the same modeling problems. The latter approach is more direct; it sidesteps single experiment excitation issues by initializing several brief experiments at "interesting" state and parameter space locations.

This type of link between systems and control theory, microscopic simulation and traditional numerical analysis may hold promise in the system level analysis of multiscale phenomena described by atomistic/stochastic simulators. A basic premise of the entire process is our assumption that we know the variables (observables of the fine scale simulation) based on which a coarse-grained, macroscopic model would close. In many cases such variables are known from previous knowledge or experimentation. For novel systems where such variables are not obvious, we need to turn to data analysis techniques for the parametrization of high-dimensional experimental or computational data. It is from such techniques (like the coordinates arising through harmonic analysis on data-based graphs (Nadler, et al., 2004)) that the variables enabling equation-free simulation will arise.

## Acknowledgments

The work of I.G.K. was partially supported through AFOSR, DARPA and an NSF/ITR grant. C.S. acknowledges partial support by the European Social Fund and National Resources - (EPEAEK II) -Pythagoras.

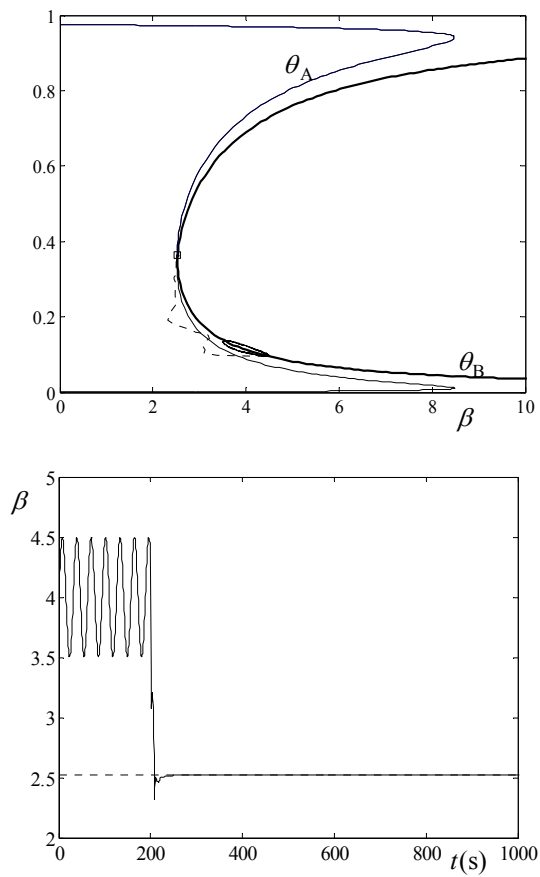

*Figure 2.: (a) Bifurcation diagram for the simplified CO oxidation mean-field model, exhibiting two turning points at β ≈ 2.5223 and β ≈ 8.4953. We overlay the phase-plane trajectory (dotted line) of the closed loop response of the SSA simulator. (b) The closed loop transient of the control variable, β. The dashed line denotes its critical value.*